\documentclass[aps,prl,reprint,superscriptaddress,amsmath,amssymb]{revtex4-2}

\usepackage{xspace}

\newcommand{\oli}{\overline}
\newcommand{\oPs}{\ensuremath{\mathrm{oPs}}\xspace}
\newcommand{\Ps}{\ensuremath{\mathrm{Ps}}\xspace}
\newcommand{\pbar}{\ensuremath{\mathrm{\oli p}}\xspace} 
\newcommand{\Hbar}{\ensuremath{\mathrm{\oli H}}\xspace}
\newcommand{\Hbarplus}{\ensuremath{\mathrm{\oli H}^+}\xspace}

\newcommand{\ep}{\ensuremath{\,\mathrm{e^+}}\xspace}
\newcommand{\el}{\ensuremath{\,\mathrm{e^-}}\xspace}

\usepackage{longtable}
\usepackage{graphicx}
\usepackage{bm}
\usepackage{orcidlink}

\usepackage{fp}
\newcount\boxheight
\newcount\boxwidth

\newcommand\testaspectone[1]{%
	\setbox0=\hbox{#1}%
	\boxheight=\ht0\relax%
	\boxwidth=\wd0\relax%
	\FPdiv\theaspect{\the\boxheight}{\the\boxwidth}%
	\FPmul\theaspect{\theaspect}{150}%
	\FPadd\theaspect{\theaspect}{20}%
	\FPeval\theaspect{round(\theaspect:1)}
	\copy0%
}

\newcommand\testaspecttwo[1]{%
	\setbox0=\hbox{#1}%
	\boxheight=\ht0\relax%
	\boxwidth=\wd0\relax%
	\FPdiv\theaspect{\the\boxheight}{\the\boxwidth}%
	\FPmul\theaspect{\theaspect}{600}%
	\FPadd\theaspect{\theaspect}{40}%
	\FPeval\theaspect{round(\theaspect:1)}
	\copy0%
}

\usepackage{hyperref}

\begin{document}


\title{
First measurement of the antihydrogen production cross section through the charge-exchange reaction of low-energy antiprotons with orthopositronium
}

\author{P.~Adrich\orcidlink{0000-0002-7019-5451}}
\affiliation{National Centre for Nuclear Research (NCBJ), ul. Andrzeja Soltana 7, 05-400 Otwock, Swierk, Poland}

\author{I.~Belosevic\orcidlink{0000-0002-8001-8889}}
\email[]{ivana.belosevic@cern.ch}
\affiliation{IRFU, CEA, Université Paris-Saclay, F-91191 Gif-sur-Yvette, France}

\author{M.~Chung\orcidlink{0000-0001-7014-4120}}
\affiliation{Pohang University of Science and Technology (POSTECH), Pohang, Republic of Korea}

\author{P.~Clad\'e}
\affiliation{Laboratoire Kastler Brossel, Sorbonne Universit\'e, CNRS, ENS-Universit\'e PSL, Coll\`ege de France, Campus Pierre et Marie Curie, 4, Place Jussieu, 75005, Paris, France}

\author{P.~Comini\orcidlink{0000-0002-6373-4752}}
\affiliation{IRFU, CEA, Université Paris-Saclay, F-91191 Gif-sur-Yvette, France}

\author{P.~Crivelli\orcidlink{0000-0001-5430-9394}}
\affiliation{Institute for Particle Physics and Astrophysics, ETH Zurich, 8093 Zurich, Switzerland}

\author{P.~Debu\orcidlink{0000-0003-2988-5052}}
\affiliation{IRFU, CEA, Université Paris-Saclay, F-91191 Gif-sur-Yvette, France}

\author{A.~Douillet}
\affiliation{Laboratoire Kastler Brossel, Sorbonne Universit\'e, CNRS, ENS-Universit\'e PSL, Coll\`ege de France, Campus Pierre et Marie Curie, 4, Place Jussieu, 75005, Paris, France}
\affiliation{Universit\'e d’Evry-Val d’Essonne, Universit\'e Paris-Saclay, Boulevard Fran\c{c}ois Mitterand, 91000 Evry, France}

\author{S.~Geffroy\orcidlink{0009-0009-6940-8120}}
\affiliation{Université Paris-Saclay, CNRS/IN2P3, IJCLab, Orsay, France}

\author{S.~Guellati-Khelifa}
\affiliation{Laboratoire Kastler Brossel, Sorbonne Universit\'e, CNRS, ENS-Universit\'e PSL, Coll\`ege de France, Campus Pierre et Marie Curie, 4, Place Jussieu, 75005, Paris, France}
\affiliation{Conservatoire National des Arts et M\'etiers, 292 rue Saint Martin, 75003 Paris, France}

\author{P.~Guichard\orcidlink{0009-0003-8568-5327}}
\affiliation{Universit\'e de Strasbourg, CNRS, IPCMS, UMR 7504, F-67000 Strasbourg, France}

\author{P.-A.~Hervieux\orcidlink{0000-0002-4965-9709}}
\affiliation{Universit\'e de Strasbourg, CNRS, IPCMS, UMR 7504, F-67000 Strasbourg, France}

\author{L.~Hilico\orcidlink{0000-0002-8916-1294}}
\affiliation{Laboratoire Kastler Brossel, Sorbonne Universit\'e, CNRS, ENS-Universit\'e PSL, Coll\`ege de France, Campus Pierre et Marie Curie, 4, Place Jussieu, 75005, Paris, France}
\affiliation{Universit\'e d’Evry-Val d’Essonne, Universit\'e Paris-Saclay, Boulevard Fran\c{c}ois Mitterand, 91000 Evry, France}

\author{P.~Indelicato\orcidlink{0000-0003-4668-8958}}
\affiliation{Laboratoire Kastler Brossel, Sorbonne Universit\'e, CNRS, ENS-Universit\'e PSL, Coll\`ege de France, Campus Pierre et Marie Curie, 4, Place Jussieu, 75005, Paris, France}

\author{S.~Jonsell\orcidlink{0000-0003-4969-1714}}
\affiliation{Department of Physics, Stockholm University, Stockholm, Sweden}

\author{J.-P.~Karr}
\affiliation{Laboratoire Kastler Brossel, Sorbonne Universit\'e, CNRS, ENS-Universit\'e PSL, Coll\`ege de France, Campus Pierre et Marie Curie, 4, Place Jussieu, 75005, Paris, France}
\affiliation{Universit\'e d’Evry-Val d’Essonne, Universit\'e Paris-Saclay, Boulevard Fran\c{c}ois Mitterand, 91000 Evry, France}

\author{B.~Kim\orcidlink{0000-0003-0958-5503}}
\affiliation{Center for Underground Physics, Institute for Basic Science, Daejeon, Korea}

\author{S.~Kim\orcidlink{0000-0002-0013-0775}}
\affiliation{Department of Physics and Astronomy, Seoul National University, Seoul, Korea}

\author{E.-S.~Kim\orcidlink{0000-0001-5603-6764}}
\affiliation{Department of Accelerator Science, Korea University Sejong Campus, Sejong, Korea}

\author{N.~Kuroda\orcidlink{0000-0003-2727-790X}}
\affiliation{Institute of Physics, University of Tokyo, Tokyo, Japan}

\author{B.~Lee\orcidlink{0000-0002-7293-5142}}
\affiliation{Department of Physics and Astronomy, Seoul National University, Seoul, Korea}
\affiliation{Present address: QUANTUM, Institut für Physik, Johannes Gutenberg Universität, Mainz, Germany}

\author{L.~Liszkay\orcidlink{0000-0003-4371-4380}}
\affiliation{IRFU, CEA, Université Paris-Saclay, F-91191 Gif-sur-Yvette, France}

\author{D.~Lunney\orcidlink{0000-0002-3227-305X}}
\affiliation{Université Paris-Saclay, CNRS/IN2P3, IJCLab, Orsay, France}

\author{G.~Manfredi\orcidlink{0000-0002-5214-8707}}
\affiliation{Universit\'e de Strasbourg, CNRS, IPCMS, UMR 7504, F-67000 Strasbourg, France}

\author{B.~Mansoulié\orcidlink{0000-0001-5945-5518}}
\affiliation{IRFU, CEA, Université Paris-Saclay, F-91191 Gif-sur-Yvette, France}
\affiliation{Department of Physics, Faculty of Science and Engineering, Swansea University, Swansea SA2 8PP, United Kingdom}

\author{V.~Martimort}
\affiliation{Laboratoire Kastler Brossel, Sorbonne Universit\'e, CNRS, ENS-Universit\'e PSL, Coll\`ege de France, Campus Pierre et Marie Curie, 4, Place Jussieu, 75005, Paris, France}

\author{M.~Matusiak\orcidlink{0000-0002-8239-6971}}
\affiliation{National Centre for Nuclear Research (NCBJ), ul. Andrzeja Soltana 7, 05-400 Otwock, Swierk, Poland}

\author{V.~Nesvizhevsky\orcidlink{0000-0002-5364-0197}}
\affiliation{Institut Max von Laue - Paul Langevin (ILL), Grenoble, France}

\author{F.~Nez\orcidlink{0000-0002-3478-7521}}
\affiliation{Laboratoire Kastler Brossel, Sorbonne Universit\'e, CNRS, ENS-Universit\'e PSL, Coll\`ege de France, Campus Pierre et Marie Curie, 4, Place Jussieu, 75005, Paris, France}

\author{K.~Park\orcidlink{0000-0002-6013-0259}}
\affiliation{Department of Physics and Astronomy, Seoul National University, Seoul, Korea}

\author{N.~Paul}
\affiliation{Laboratoire Kastler Brossel, Sorbonne Universit\'e, CNRS, ENS-Universit\'e PSL, Coll\`ege de France, Campus Pierre et Marie Curie, 4, Place Jussieu, 75005, Paris, France}

\author{E.~Perez\orcidlink{0000-0002-7359-9689}}
\affiliation{CERN, EP Department, 1 Esplanade des Particules, 1217 Meyrin, Switzerland}

\author{P.~Pérez\orcidlink{0000-0003-1407-1582}}
\affiliation{IRFU, CEA, Université Paris-Saclay, F-91191 Gif-sur-Yvette, France}
\affiliation{Department of Physics, Faculty of Science and Engineering, Swansea University, Swansea SA2 8PP, United Kingdom}

\author{C.~Regenfus\orcidlink{0000-0001-9656-3104}}
\affiliation{Institute for Particle Physics and Astrophysics, ETH Zurich, 8093 Zurich, Switzerland}

\author{C.~Roumegou\orcidlink{0009-0006-0343-9256}}
\affiliation{Université Paris-Saclay, CNRS/IN2P3, IJCLab, Orsay, France}

\author{J.-Y.~Rouss\'e\orcidlink{0000-0000-0000-0000}}
\author{F.~Schmidt-Kaler\orcidlink{0000-0002-5697-2568}}
\affiliation{QUANTUM, Institut für Physik, Johannes Gutenberg Universität, Mainz, Germany}

\author{K.~Szymczyk\orcidlink{0000-0001-9159-485X}}
\affiliation{National Centre for Nuclear Research (NCBJ), ul. Andrzeja Soltana 7, 05-400 Otwock, Swierk, Poland}

\author{T.~A.~Tanaka\orcidlink{0009-0003-8306-5546}}
\affiliation{Institute of Physics, University of Tokyo, Tokyo, Japan}
\affiliation{Present address: National Metrology Institute of Japan (NMIJ), National Institute of Advanced Industrial Science and Technology (AIST), Tsukuba, Japan}

\author{B.~Tuchming\orcidlink{0000-0002-1356-0723}}
\affiliation{IRFU, CEA, Université Paris-Saclay, F-91191 Gif-sur-Yvette, France}

\author{D.~P.~van~der~Werf\orcidlink{0000-0001-5436-5214}}
\affiliation{Department of Physics, Faculty of Science and Engineering, Swansea University, Swansea SA2 8PP, United Kingdom}

\author{D.~Won\orcidlink{0009-0005-5557-7709}}
\affiliation{Department of Physics and Astronomy, Seoul National University, Seoul, Korea}

\author{S.~Wronka\orcidlink{0000-0003-3277-138X}}
\affiliation{National Centre for Nuclear Research (NCBJ), ul. Andrzeja Soltana 7, 05-400 Otwock, Swierk, Poland}

\author{P.~Yzombard\orcidlink{0000-0002-0864-181X}}
\affiliation{Laboratoire Kastler Brossel, Sorbonne Universit\'e, CNRS, ENS-Universit\'e PSL, Coll\`ege de France, Campus Pierre et Marie Curie, 4, Place Jussieu, 75005, Paris, France}

\collaboration{GBAR Collaboration}

             
\begin{abstract}

The GBAR experiment has measured the formation rate of antihydrogen from antiproton impact on a positronium cloud for antiproton kinetic energies of 4 and 6 keV.  
This is the first charge-exchange cross section measurement performed using antiproton beams, which are provided by the AD-ELENA facility at CERN.
The measured cross section values are $(14.1 \pm 1.3 \mathrm{(stat)} ^{+2.2}_{-1.4} \mathrm{(sys)}) \times 10^{-16}$~cm$^2$ at 6.2~keV energy and $(8.7 \pm 2.4 \mathrm{(stat)} ^{+1.5}_{-0.09} \mathrm{(sys)}) \times 10^{-16}$~cm$^2$ at 4.15~keV energy, and agree with recent theoretical three-body calculations for antihydrogen formation.
These measurements are an important input for experiments with antimatter, e.g the planned measurements of antihydrogen gravitational acceleration in the GBAR collaboration.

\end{abstract}

\maketitle

\section{Introduction}

Experiments with antimatter at low energy are important for 
testing the Standard Model of Particle Physics, especially when probing CPT symmetry or exploring the role of gravity~\cite{safronova_search_2018,villata_cpt_2011}.  Antihydrogen offers excellent opportunities for such tests, which are pursued by several experiments~\cite{malbrunot_asacusa_2018,baker_precision_2025,amsler_pulsed_2021,anderson_observation_2023} at CERN's AD-ELENA facility~\cite{carli_elena_2022}.

As outlined in the review article~\cite{doser_antiprotonic_2022}, antihydrogen (\Hbar) can be produced through three processes:  three-body recombination, radiative recombination, and charge exchange.
Three-body recombination, whereby a scattered positron ($\mathrm{e^+}$) is captured by an antiproton (\pbar), while another \ep carries away the extra energy, is used by ASACUSA~\cite{malbrunot_asacusa_2018} and ALPHA~\cite{baker_precision_2025,anderson_observation_2023} experiments.

In contrast, the GBAR experiment (Gravitational Behaviour of Antimatter at Rest), which is designed to investigate the weak equivalence principle by measuring the free-fall acceleration of antihydrogen in the Earth's gravitational field, takes a different approach to producing ultracold antihydrogen atoms: first, antihydrogen ions (\Hbarplus) are planned to be produced through two successive charge-exchange reactions with positronium atoms (Ps)~\cite{perez_gbar_2015}:

\begin{eqnarray}
\pbar + \Ps &\rightarrow \Hbar + \el \label{eq1} \\ 
\Hbar + \Ps &\rightarrow \Hbarplus + \el \label{eq2}
\end{eqnarray}

The antihydrogen ions are then sympathetically cooled using
Be$^+$ ions~\cite{sillitoe_bartexth_2017, sillitoe_production_2017}. Afterwards, the extra positron is photo-detached from cooled antihydrogen ions to obtain antihydrogen atoms with temperatures on the order of 10 $\mu$K.

Charge exchange by Reaction \eqref{eq1} has been used by ATRAP~\cite{atrap_collaboration_first_2004} and AEGIS~\cite{amsler_pulsed_2021} to produce neutral antihydrogen from positronium excited to Rydberg states. GBAR has also demonstrated the production of antihydrogen through charge exchange, but with ground state ortho-positronium (oPs)~\cite{adrich_production_2023}. 
At that time, the insufficient control of detector acceptance made a measurement of the production cross section impossible.

The positronium charge-exchange reaction cross section was measured for hydrogen between 11 and 16 keV by Merrison et al. \cite{merrison_hydrogen_1997} in 1997.
In this Letter, almost 30 years later, we report the measurement of the antimatter counterpart reaction at 4 and 6~keV, 
produced by the  GBAR experiment at CERN’s AD/ELENA facility, providing first quantitative benchmark of modern three-body calculations in the antihydrogen channel. These results are crucial for the GBAR experimental scheme, yielding input for the rates at which the GBAR setup may generate antihydrogen ions.

\begin{figure*}[ht!]
\testaspecttwo{
\includegraphics[width=1\linewidth]{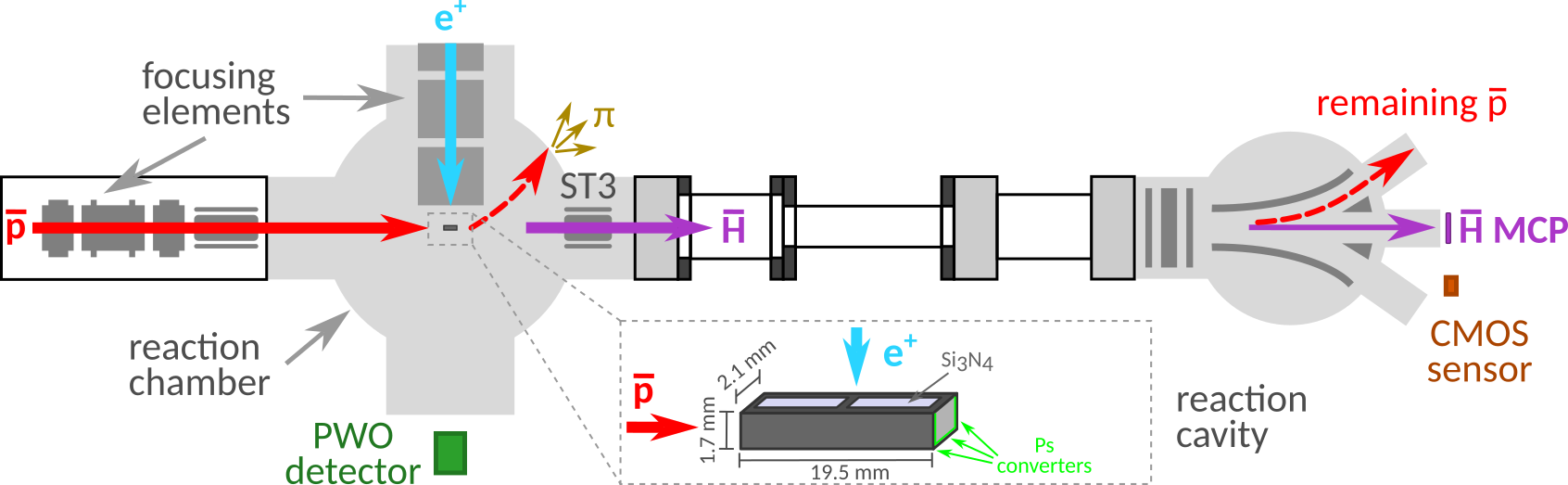}
}
\caption{\label{fig:gbar} Schematic illustration of the GBAR set-up and production scheme of antihydrogen.  
Positrons produced using a linac-based source are accumulated in a set of two traps (not shown) and directed to the reaction chamber, where they are converted into orthopositronium. Antiprotons from ELENA are decelerated with a drift tube and injected into a Penning-Malmberg trap (not shown). After electron cooling and rotating-wall compression, the \pbar pulse extracted from the trap is transported to the Ps target cavity (inset close-up). The antiprotons that did not interact with Ps are deflected (see red dashed lines).  The \Hbar, \Ps and \pbar are counted with MCPs, scintillators, and a CMOS detector, respectively. 
}
\end{figure*}

\section{Experiment}

The \Hbar production and detection scheme is shown in Fig.~\ref{fig:gbar}, and is described in detail in Ref.~\cite{adrich_production_2023}.  For this work, several upgrades were implemented that resulted in an order of magnitude improvement in \Hbar production rate, notably the addition of a Penning-Malmberg trap for cooling and accumulation of antiprotons~\cite{lee_record_2026} and the implementation of a cavity-based positronium converter to enhance the oPs cloud density.

For positron production, we use a linac-based positron source coupled with a tungsten-mesh moderator~\cite{charlton_positron_2021}. The low energy positrons are transported to a modified buffer gas trap (BGT), where they are captured and remoderated using a combination of CO$_2$ cooling gas and SiC remoderator~\cite{liszkay_news_2025}. 

The cooled positron bunches are then sent to a high-field Penning-Malmberg trap (HFT), where about 600 positron bunches from the BGT are accumulated between successive \pbar shots (every two minutes), resulting in approximately $3 \times  10^8$ trapped positrons in the HFT.%
The positrons from the HFT are re-accelerated to 4.3~keV energy and transported to the reaction chamber using a series of electrostatic elements. The reaction chamber houses a movable target holder that allows us to change between different targets 
(in particular, a tube-shaped cavity with an inner layer of Ps converter, and other targets used for positronium conversion efficiency reference). 
A box made of 3 mm-thick soft iron plates shields the reaction chamber area from the fringe field of the HFT.
A grid with 92~\% transparency is installed 14~mm upstream of the reaction cavity to shield it from the high-voltage Einzel lenses of the positron transport beamline. The number of positrons arriving in the reaction chamber is measured periodically from their deposited charge on 
one of the targets,
a large-area plate in the plane of the target holder, polarized at +100~V, 
and connected to a charge-sensitive preamplifier (CSP). 
Typically, $10^8$ \ep with 6.5~ns (RMS) time spread reach the reaction chamber, with the main losses occurring at the transition from the high to low magnetic field regions.

Finally, the positron bunch is focused into the reaction cavity of about $1.7 \times 2.1 \times 19.5 $~mm$^3$ size (see inset in Fig.~\ref{fig:gbar}) through a 30~nm-thick $\mathrm{Si_3N_4}$ window. About 30\% of positrons arriving in the reaction chamber enter the cavity (efficiency given by the geometrical acceptance of the window) and stop in the walls of the reaction cavity, which are coated with a mesoporous SiO$_2$ material. 
At 4.3 keV, almost 30\% of the positrons stopped in the mesoporous layer are converted into oPs, 
and 70\% of these are emitted into vacuum at 50~meV thermal energy~\cite{crivelli_measurement_2010}, creating a semi-confined cloud of about $6 \times 10^6$ ground-state oPs atoms per positron bunch injection.

The positronium yields are regularly monitored using a 
PbWO$_4$ detector (PWO in the following), placed close to the reaction chamber (see Fig.~\ref{fig:gbar})~\cite{kim_development_2020}. Gamma rays emitted from the positron and Ps annihilations produce scintillation light in the PbWO$_4$ crystal, which is then collected using a fast photomultiplier.

A 100~keV \pbar beam is provided by ELENA~\cite{carli_elena_2022}, typically with $12\times10^6$ \pbar and 40~ns width ($\sigma$) in a single shot. The beam is decelerated to 3~keV using a pulsed drift tube~\cite{husson_pulsed_2021}, and subsequently transported to a Penning-Malmberg trap (antiproton trap)~\cite{lee_record_2026}, where it is captured and cooled. The antiprotons are co-trapped with 3~$\times 10^8$ electrons, which cool the \pbar plasma. At the same time, a rotating-wall electric field (strong drive)~\cite{danielson_radial_2006} is applied to compress 
the \pbar plasma. After the cooling, the antiprotons are re-accelerated to either 4 or 6~keV using a pulsed drift tube accelerator-buncher.  The bunching at the trap ejection is set to time-focus the \pbar bunch at the position of the cavity, with a typical bunch width ($\sigma$) of 60~ns.
The total injection-trapping-ejection efficiency per \pbar shot is 40-50\%, resulting in bunches containing about $5-6 \times 10^6$ antiprotons.

The antiprotons are then transported downstream and focused into the reaction cavity using a series of electrostatic beamline elements. 
The antiprotons interact with the oPs cloud that is semi-confined in the reaction cavity.
The delay between the antiproton pulse and the positron pulse is adjusted to get the optimal overlap between antiprotons and oPs atoms. The \Hbar atoms produced in the charge-exchange reaction are detected in the detection micro-channel plate detector (MCP) 1.6~m downstream. The remaining antiprotons are deflected toward the walls of the reaction chamber using an electrostatic deflector placed immediately after the reaction cavity.
Care was taken to compensate for the HFT residual magnetic field after the cavity with an electrostatic steerer (ST3 in Fig.~\ref{fig:gbar}) so that the antihydrogen beam and 
the antiproton beam, when the cavity deflector is off, arrive at the same position at the detection MCP, thus allowing us to use the undeflected \pbar beam to monitor the acceptance of the \Hbar beam on the detection MCP. The undeflected \pbar beam was also used to monitor the number of antiprotons passing through the cavity with a CMOS sensor placed downstream from the cavity, which detects products of the \pbar annihilating on the front face of the detection MCP~\cite{regenfus_monitoring_2026}.
The beamline upstream of the cavity was optimized to maximize the number of antiprotons passing through the cavity, while ensuring 100\% acceptance on the detection MCP.

We recorded two types of runs: mixing runs (where both \pbar and \ep were present) and background runs (with only \pbar present). The background comes from secondary particles produced in antiproton annihilations upstream of the detection MCP. Positrons were previously found to produce no significant background~\cite{adrich_production_2023} in the detection MCP. The total accumulated statistics is 3893 mixing events and 3409 background events for the 6~keV run, and 2007 mixing events and 1309 background events for the 4~keV run.

\section{Results and Discussion}

The cross-section of reaction~\eqref{eq1} is given by:

\begin{equation}
   \sigma (\oli v_{\pbar}) = \frac{N_{\Hbar}}{N_{\pbar\mathrm{,tot}} \cdot N_{\mathrm{oPs}} } \cdot \frac{A_{\mathrm{cavity}}}{I_{\mathrm{overlap}}} ,
   \label{eq:cs}
\end{equation}

where $N_{\pbar\mathrm{,tot}}$ is the total number of \pbar in all the mixing spills, $N_{\mathrm{oPs}}$ is the average oPs number over all mixing shots,  $N_{\Hbar}$ is the number of produced antihydrogen atoms, and $A_{\mathrm{cavity}} = 3.48 \pm 0.10$~mm$^2$ is the transverse area of the cavity (perpendicular to the direction of \pbar propagation). The dimensionless overlap factor $I_{\mathrm{overlap}}$ describes geometrical and temporal overlap between oPs and \pbar. In the simple case where the oPs density is taken as homogeneous and limited to the cavity volume, $I_{\mathrm{overlap}}$ can be written as follows:

\begin{equation}
    I_{\mathrm{overlap}} \equiv  \frac{\oli v_{\pbar}}{L_{\mathrm{cavity}}} \int F_{\pbar} (t) \cdot F_{\mathrm{oPs}} (t) \cdot dt,
    \label{eq:overlap_function}
\end{equation}

where $\oli v_{\pbar}$ is the mean \pbar velocity,  $L_{\mathrm{cavity}}$ is the length of the reaction cavity and $F_{\pbar} (t)$ and $F_{\oPs} (t)$ are \pbar and oPs temporal distributions, respectively, normalized to unity. Since the mean energy of the emitted \oPs is only 50~meV, the relative velocity between \pbar and \oPs is essentially given by the \pbar velocity.
To extract the cross section, we need to determine absolute numbers for \pbar, \Hbar, and oPs, and the overlap between \pbar and oPs pulses. Each of these ingredients is discussed in the following.

\begin{figure*}
    \centering
    \testaspecttwo{
    \includegraphics[width=0.454\linewidth]{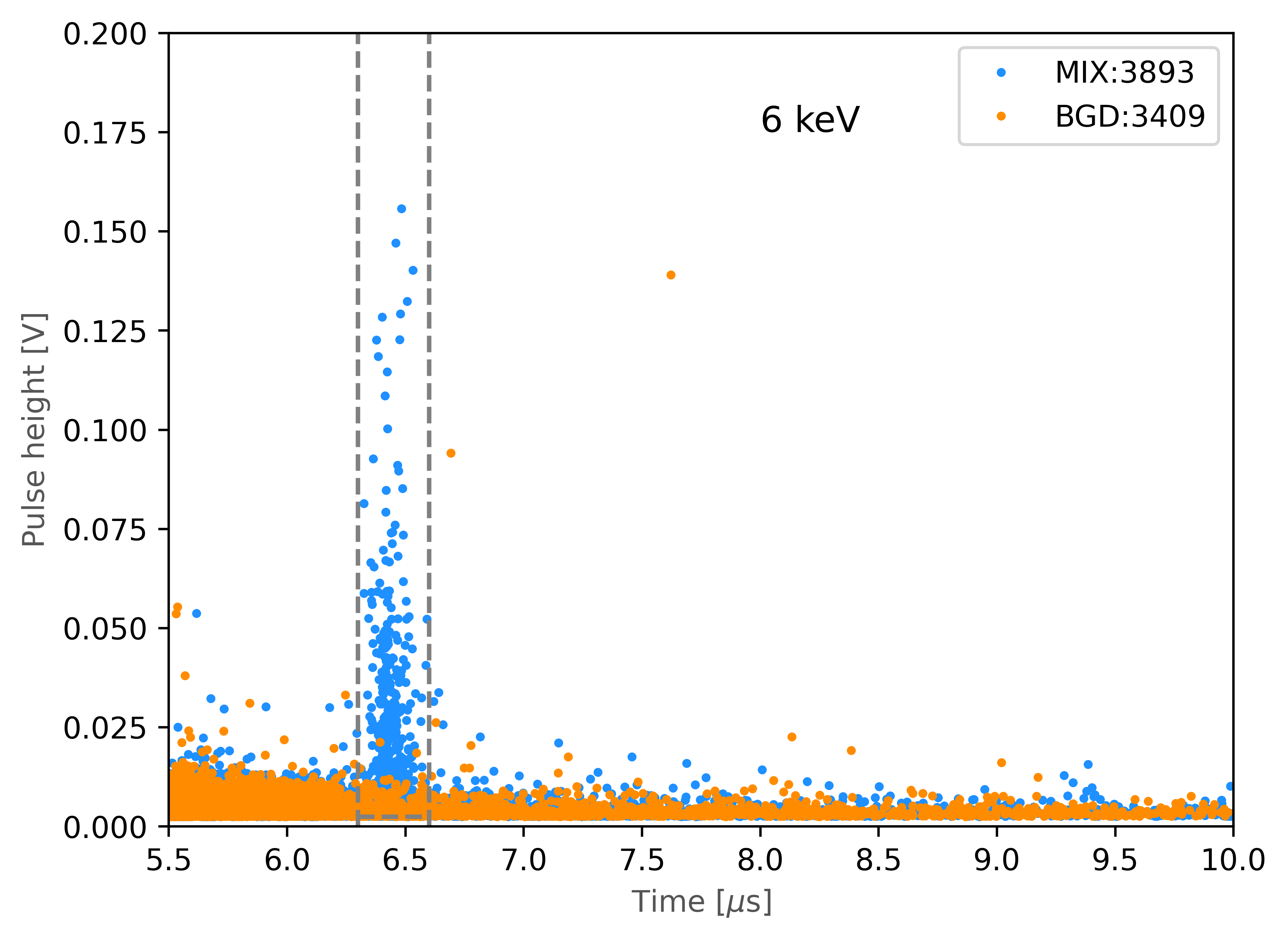}
    \includegraphics[width=0.454\linewidth]{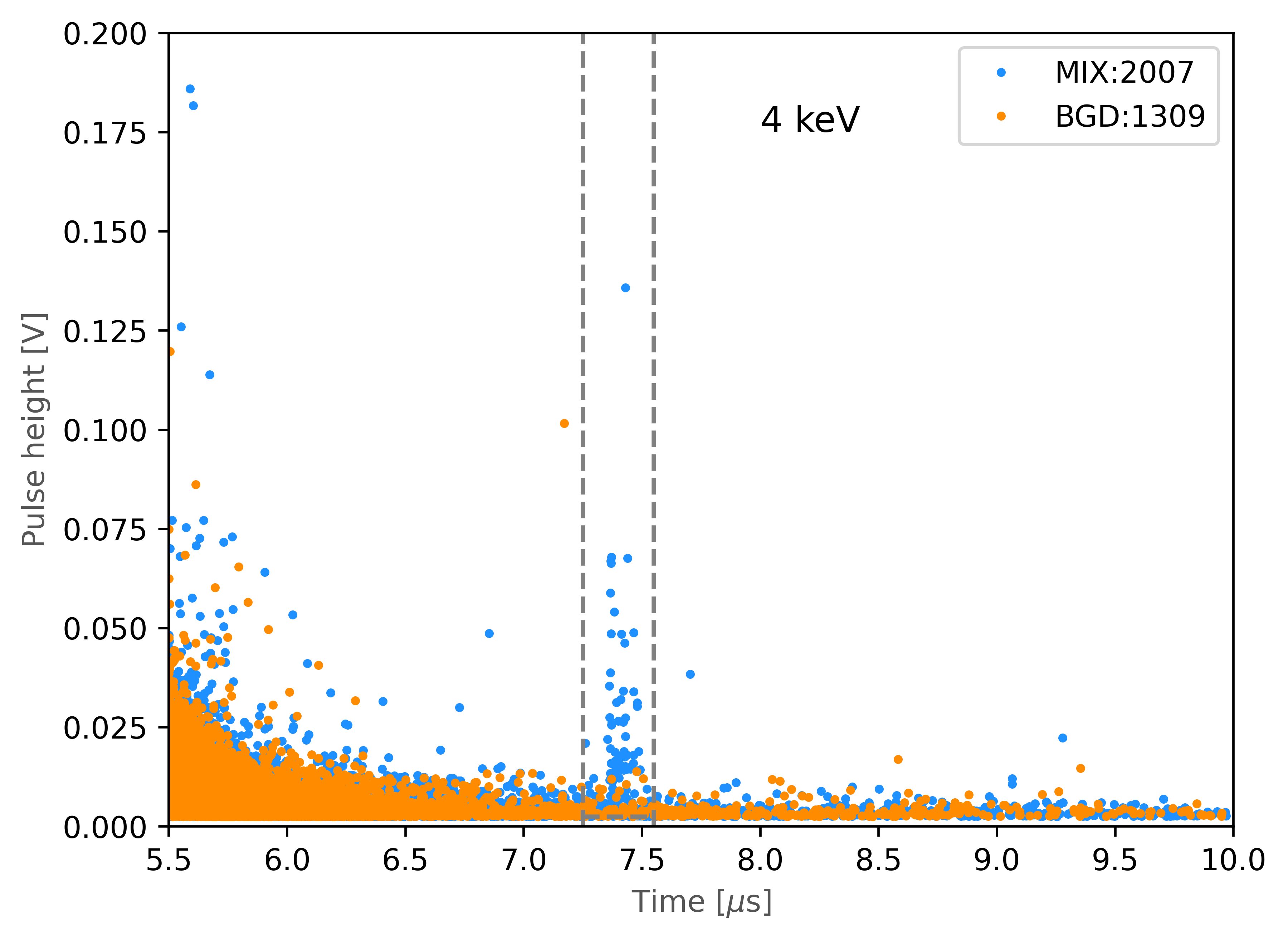}
    }
    \caption{Pulse height versus pulse time distribution of recorded MCP electrical signal for mixing (blue) and background (orange) runs for 6~keV (left) and 4~keV (right) datasets. The dashed lines indicate signal counting windows.}
    \label{fig:el_signal_analysis}
\end{figure*}

\paragraph{Antihydrogen atoms}
Antihydrogen atoms created in the reaction chamber travel downstream and are detected by the detection MCP. 
Fig.~\ref{fig:el_signal_analysis} shows the peak height of the electrical signal versus the peak time for mixing and background runs. 
A clear excess of large amplitude signals is seen in the mixing runs for both 4~keV and 6~keV datasets in the expected time windows (given by time of flight of \pbar/\Hbar beam).
The signal region is defined as pulses having amplitudes greater than 2 mV in the expected signal time window (6.3-6.6~$\mathrm{\mu}$s for 6~keV and  7.25-7.55~$\mathrm{\mu}$s for 4~keV). The signal yield is obtained as the number of peaks in the mixing runs minus that in the background runs normalized by the relative number of shots. This gives $379 \pm 35$ detected \Hbar atoms for the 6~keV dataset, and $69 \pm 19$ detected \Hbar atoms for the 4~keV dataset.

To obtain the actual \Hbar yield, these numbers are corrected for several
effects: backscattering of \Hbar from the MCP surface,
and MCP detection efficiency. 
The MCP detection efficiency is estimated using a GEANT4 simulation~\cite{park_mcp_nodate} to be $\varepsilon_{\mathrm{det, MCP}} = 96.78 \pm 0.01~\mathrm{(stat)} \pm 0.31 ~\mathrm{(sys)}$\%. 
\Hbar atoms that backscatter from the surface produce no signal in the MCP. We studied this effect by impinging 4 keV, 6 keV, and undecelerated 100 keV antiproton beams from ELENA on a MCP immediately after the pulsed drift tube decelerator. The assumption is that antihydrogen atoms backscatter in the same way as antiprotons since antihydrogen atoms are quickly stripped of the positron when they enter the surface of the MCP. A CMOS sensor was placed at different distances R from the beam axis to detect the products of the \pbar annihilations. As the backscattered antiprotons would annihilate away from the MCP annihilation plane (in the vacuum chamber wall, on average about 5~cm upstream of the MCP plane), this would cause a departure from the simple 1/R$^2$ scaling of the CMOS detection acceptance given by the solid angle of the detector. By fitting these measurements using templates for various backscattering probabilities generated by GEANT4~\cite{Agostinelli2003}, a backscattering probability $f_{\mathrm{BS}}$
compatible with zero, with an upper limit (1$\sigma$) of 11.5\% and 14\% respectively at 6 and 4 keV was obtained
~\cite{park_determination_2026}. 

After applying the corrections for the MCP efficiency and \Hbar backscattering, we obtain 392.0 $\pm$ 36.1(stat)~$^{+45.1}_{-1.3}$(sys) produced \Hbar atoms in the 6~keV dataset, and 71.4 $\pm$ 19.4(stat)~$^{+10.0}_{-0.2}$(sys) produced \Hbar atoms in the 4~keV dataset.

\paragraph{Antiproton number}
The number of antiprotons during the data-taking is measured regularly using a CMOS sensor placed at 34~cm distance from the center of the \Hbar detection MCP. To extract the absolute number of antiprotons involved in the reaction, the expected number of tracks $N_{\mathrm{clusters}}$ detected in the CMOS sensor per antiproton annihilation has been predicted using GEANT4 simulations and independently measured ELENA intensities~\cite{park_determination_2026}. Using this calibration, we obtain the average number of $(7.96 \pm 0.06 \mathrm{(stat)} \pm 0.67 \mathrm{(sys)}) \times 10^{5}$ \pbar passing through the reaction cavity during the 6~keV run, and $(4.21 \pm 0.05\mathrm{(stat)} \pm 0.35 \mathrm{(sys)}) \times 10^{5}$ \pbar during the 4~keV run. The main contributions to the 8.4\% overall uncertainty come from 4\% uncertainty on the intensity measured by ELENA monitors used for calibration and 5\% uncertainty accounting for a potential difference between the backscattering fractions on the detection MCP and on the MCP used for the calibration measurements~\cite{park_determination_2026}.

\paragraph{oPs number}

The total number of oPs produced is given by $N_{\mathrm{oPs}} = N_{e^+} \cdot f_{\mathrm{oPs}}$, where $ N_{e^+}$ is the number of positrons arriving in the reaction chamber and $f_{\mathrm{oPs}}$ is the fraction of positrons that get converted to oPs.

The number of positrons reaching the target plane and measured with the CSP is corrected to account for positron backscattering from the target (13\% on silicon according to GEANT4 simulation), and also corrected for the contribution of the secondary electrons $k_{SE}$ that are created by positron impact on the grid upstream the reaction cavity, estimated to be on the order of +2\% using COMSOL\textsuperscript{\textregistered} simulation.

A systematic uncertainty of 3\% is assigned to the charge-sensitive preamplifier calibration, based on the comparison of the manufacturer's calibration with our own calibration using charge injection via a precision test capacitor~\cite{comini_note_2025}. The obtained average number of \ep arriving in the reaction chamber during the 6~keV measurement is 
$(1.059 \pm 0.003 \mathrm{(stat)} \pm 0.032 \mathrm{(sys)}) \times 10^8$ and 
$(1.116 \pm 0.005 \mathrm{(stat)} \pm 0.034\mathrm{(sys)}) \times 10^8$ for the 4~keV measurement.

The oPs fraction is extracted by applying the single-shot positronium annihilation lifetime spectroscopy technique (SSPALS)~\cite{deller_sspals_2019}, fitting the PWO data with a template that is a sum of a prompt annihilation peak (positron and para-Ps annihilations) weighted by $(1-f_{\mathrm{oPs,fit}})$ and the oPs decay weighted by $f_{\mathrm{oPs,fit}}$, taking into account both the decay inside the pores of the converter with a short lifetime $\tau_{p}$, and the decay in vacuum, for the oPs atoms that escaped the pores at a rate $\kappa_v$, with the lifetime $\tau_{v} = 142$~ns~\cite{comini_note_2025}. Both $\tau_{p}$ and $\kappa_v$ are determined from an independent measurement~\cite{crivelli_measurement_2010}, with values given in Table~\ref{tab:cs_corrections}.
The template is generated using PWO scintillator data taken on a special target, a reference cavity where the Ps conversion is suppressed~\cite{comini_note_2025}.

The typical fitted \oPs fractions range from 6.8\% to 7.6\%.
The fitted \oPs fraction $f_{\mathrm{oPs,fit}}$ needs to be corrected for the positron transport losses upstream of the reaction cavity, since these positrons still contribute to the prompt annihilation peak, but do not contribute to the \oPs production, thus resulting in a lower apparent \oPs fraction. This number is determined by comparing the \oPs fraction measured on a large flat Ps-converter target ($19\times19$~mm$^2$), which fits the full positron beamspot, for different sizes of positron bunches: for $9 \times 10^7$ \ep (the same number as during the cross section measurements), and for a smaller bunch of $3 \times 10^7$ \ep (in which case, the transport losses within the acceptance region of the PWO detector are negligible). The measured \oPs fraction with the smaller positron bunch is compatible with independently measured oPs conversion efficiency for the same type of converter using the PALS technique~\cite{crivelli_measurement_2010}. The difference in the \oPs fraction between the large and small positron bunch implantation is around 22\%. We attribute this difference to the transport losses upstream of the reaction chamber. 
A loss of 5\% in the grid located in front of the reaction cavity is also corrected for, leading to a final correction $k_{\mathrm{loss}}$ presented in Table~\ref{tab:cs_corrections}.

By applying this correction, we obtain the average total oPs number $N_{\mathrm{oPs}} = N_{e^+} \cdot f_{\mathrm{\oPs,fit}} \cdot k_{\mathrm{loss}}$: $(9.53 \pm 0.09 \mathrm{(stat)} \pm 0.35 \mathrm{(sys)}) \times 10^6$ for the 6~keV dataset and $(10.26 \pm 0.12\mathrm{(stat)} \pm 0.49 \mathrm{(sys)}) \times 10^6$ for the 4~keV dataset. The main contributions to the systematic uncertainty are 2-3\%  uncertainty due the systematic uncertainty on the factor $k_{\mathrm{loss}}$, and 3\% due to CSP calibration uncertainty, see table~\ref{tab:cs_corrections}.
The separation between the oPs population in the pores and the oPs population in vacuum, emitted in the cavity volume,
is accounted for in the overlap factor, as explained in the following.

\paragraph{Overlap factor}
The two inputs that are needed to determine the overlap factor, \pbar velocity, and the cavity length, are presented in Table~\ref{tab:cs_corrections}. The mean \pbar velocity is obtained from time of flight (ToF) measurements at different MCPs along the antiproton beamline, giving  $\overline v_{\mathrm{\pbar,\,6~keV}} = 1.0889 \pm 0.0005 \mathrm{(stat)} \pm  0.0044 \mathrm{(sys)}\, \mathrm{mm/ns}$ and $v_{\mathrm{\pbar,\,4~keV}} = 0.8930 \pm 0.0018 \mathrm{(stat)} \pm 0.0036 \mathrm{(sys)}\, \mathrm{mm/ns}$, corresponding to 6.2~keV and 4.15~keV mean \pbar energy, respectively.
The integral in the overlap factor (Eq.~\eqref{eq:overlap_function}) is calculated numerically using the experimental \pbar pulse shape measured on the MCP in the reaction chamber $F_{\pbar}(t)$ and the
 distribution of the oPs population in vacuum $F_{\oPs} (t)$. With values for $\kappa_v$ and $\tau_p$ from Table~\ref{tab:cs_corrections} and the RMS of \ep pulse of 6.5~ns, $F_{\oPs}(t)$ peaks at about 50~ns after the \ep arrival on the converter target. 

Finally, we obtain $I_{\mathrm{overlap}} = 0.3276 \, ^{+0.0003}_{-0.0011} \mathrm{(stat)} \pm 0.010 \mathrm{(sys)}$ and $I_{\mathrm{overlap}} = 0.330 \, ^{+0.003}_{-0.005} \mathrm{(stat)} \pm 0.010 \mathrm{(sys)}$ for the 6~keV and 4~keV datasets respectively. The main contributions to the systematic uncertainty is 3\% due to uncertainties on parameters $\kappa_v$ and $\tau_p$.
The uncertainty on length of the cavity has negligible influence on the overlap factor. The statistical uncertainty comes mainly from the 6~ns jitter 
in the delay between \pbar and \oPs pulse.
This statistical uncertainty in the 4~keV case is relatively large, because the experimental delay was found to be 15~ns earlier than the optimal value, thus leading to a larger sensitivity to the delay jitter than in the 6~keV measurement.

\begin{table}[h!bt]
    \centering
    \begin{tabular}{ l|p{2cm}p{3.0cm}}
        quantity & value & rel. sys. uncertainty\\
    \toprule
        $1-f_{\mathrm{BS}}$  & $1.0$ & $^{+0}_{-11.5}\%$ ($^{+0}_{-14}\%$) \\
        $\varepsilon_{\mathrm{det, MCP}}$ &  $0.9678$ & $0.32\%$ \\
        $N_{\mathrm{clusters}}/\pbar$ annihilation & $2.4 \times 10^{-4}$ & $8.4\%$ \\
        $k_{\mathrm{SE}}$  & $1.022$ $(1.017)$& 0.3\% (0.6\%)\\
        $k_{\mathrm{loss}}$  & $1.287 (1.280)$ & 1.5\% (3.3\%) \\
        $\tau_{p}$ & 75~ns & 2.7\% \\
        $\kappa_{v}$ & 0.0348~ns$^{-1}$ & 8\% \\
        \pbar velocity & 1.09\,mm/ns& 0.4\%\\
         & (0.89~mm/ns)& (0.4\%)\\
        $A_{\mathrm{cavity}}$  & 3.48~$\mathrm{mm^2}$ & 2.8\% \\
        $L_{\mathrm{cavity}}$  & 19.5~$\mathrm{mm}$ & 2.5\% \\
        CSP calibration & 12.5~mV/pC & 3\% \\

    \end{tabular}
    \caption{Sources of systematic errors that enter into cross section determination. The numbers within parentheses are for 4~keV dataset (only where they differ from the 6~keV dataset systematics). 
    }
    \label{tab:cs_corrections}
\end{table}

\paragraph{Cross section results and discussion of systematics}

The sources of different contributions to the systematic uncertainty are summarized in Table~\ref{tab:cs_corrections}. The systematic uncertainty is ultimately dominated by the systematic uncertainties assigned to the antiproton/antihydrogen backscattering fractions, which enter both into the final number of the produced antihydrogen atoms and into the calibration of the CMOS sensor, used to determine absolute \pbar rates. 

\begin{figure}
    \centering
    \testaspectone{
    \includegraphics[width=1\linewidth]{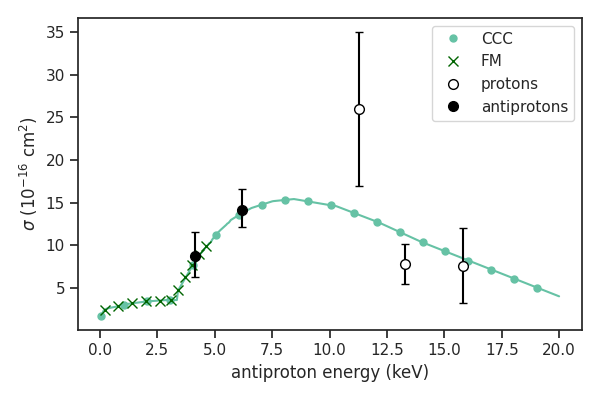}
    }
    \caption{Measured cross section for \Hbar\ formation by charge exchange between antiprotons and positronium atoms at 4 and 6~keV, compared with the only existing equivalent proton experiment \cite{merrison_hydrogen_1997} and with calculations using the Convergent Close Coupling method (CCC) \cite{rawlins_calculation_2016} and the Faddeev-Merkuriev (FM) equations \cite{valdes_ab_2018}, the latter being only available up to 4.6~keV.}
    \label{fig:cs-result}
\end{figure}

Finally, inserting the numbers above in Eq.~\eqref{eq:cs}, we obtain the cross section of reaction~\eqref{eq1} of $(14.1 \pm 1.3 \mathrm{(stat)} ^{+2.2}_{-1.4} \mathrm{(sys)}) \times 10^{-16}$~cm$^2$ at 6.2~keV energy and $(8.7 \pm 2.4 \mathrm{(stat)} ^{+1.5}_{-0.09} \mathrm{(sys)}) \times 10^{-16}$~cm$^2$ at 4.15~keV energy. These values are reported in Fig.~\ref{fig:cs-result}, where they are compared with theoretical predictions. The measurements with protons from Ref.~\cite{merrison_hydrogen_1997} are also shown in the same figure. 

Theoretically, the measured process should have the same cross section as its charge conjugate. Early calculations based on various variations of the Born approximation gave differing results~\cite{mitroy_formation_1995, comini_bar_2013, leveque-simon_antihydrogen_2023}. Considering that this is a high-energy approximation, and that the centre-of-mass energy for 6 keV (4 keV) collisions is only 6.5 eV (4.4 eV), this is not surprising. More recently, complementary calculations based on the exact non-relativistic quantum-mechanical equations, using the Convergent Close Coupling method (CCC)~\cite{rawlins_calculation_2016}, and the Faddeev-Merkuriev (FM) equations solved using a mesh of Lagrange-Laguerre basis functions~\cite{valdes_ab_2018} show excellent agreement with each other and with our measurements.

\section{Conclusion}

We report the cross-section measurement for the production of antihydrogen via charge exchange with ground-state orthopositronium, with lower uncertainties than the measurements published for the matter-analog reaction producing hydrogen. 
Comparison with two independent ab initio three-body calculations shows excellent agreement, validating modern positronium-mediated antihydrogen formation models.

Antihydrogen ion production, which is the next step of the GBAR project, requires the knowledge of the cross section of the four-body charge-exchange reaction~\eqref{eq2}.
This is presently being pursued
with the matter-analog reaction using 
the H$^{-}$ beam also provided by ELENA.
The knowledge of the rate for Reactions~\eqref{eq1} and \eqref{eq2}, in combination with the transport, the capture, and the detection efficiency determines the expected data rate of the planned free-fall experiment with ultra-cold antihydrogen. Moreover, in combination with the modelling of the experimental measurement scheme, this rate will determine the anticipated accuracy limits for testing gravity with antimatter.

\begin{acknowledgments}

We thank L. Ponce and the AD/ELENA team as well as F. Butin and the CERN EN team for
their fruitful collaboration. We also thank A. Beynel for his continued support with the alignment
of the trap. This work is supported by: JSPS KAKENHI Grant-in-Aid for Scientific Research A 20H00150 and Fostering Joint International Research A 20KK0305 (Japan), %
the Action Thématique GRAM of CNRS/INSU with INP and IN2P3 co-funded by CNES (France),%
SPHINX ANR-22-CE31-0019 (France), ESPRIT ANR-22-CE30-0028-01 (France), BESCOOL ANR/DFG ANR-13-ISO4-0002-01 (France/Germany), ANR-11-LABX-0058-NIE and ANR-17-EURE-0024 (within the Investissements d’Avenir program (ANR-10-IDEX-0002-02),%
Polish Ministry of Science and Higher Education under the program 'Support for the participation of Polish research teams in international research infrastructure projects' contract No. 2022/WK/11 (Poland),%
the Swiss National Science Foundation (Switzerland) grants 197346, 216673 and 232699 and ETH Zurich (Switzerland) grant ETH-46 17-1,%
the Swedish Research Council (VR) grants 2017-03822, 2021-04005 and 2025-04378, the German cluster of excellence PRISMA, and the following grants from Korea:
IBSR016-Y1, IBS-R016-D1, UBSI Research Fund (No. 1.220116.01) of UNIST, POSTECH Initial Settlement Support Fund, NRF-2016R1A5A1013277, NRF-RS-2022-00143178,
NRF-2021R1A2C3010989, and NRF-2016R1A6A3A11932936. The GBAR collaboration is an
International Research Network, supported by CNRS, France.

\end{acknowledgments}

\bibliography{references-new}

\end{document}